\documentclass[12pt,preprint]{aastex}

\shorttitle{SASI around a Kerr Black Hole }
\shortauthors{Hiroki Nagakura}

\begin{document}

\title{ The Standing Accretion Shock Instability
 in the Disk around the Kerr Black Hole}

\author{Hiroki Nagakura \altaffilmark{1}and Shoichi Yamada\altaffilmark{1,2}}

\altaffiltext{1}{Science and Engineering, Waseda University, 3-4-1
Okubo, Shinjuku, Tokyo 169-8555, Japan}
\altaffiltext{2}{Advanced Research Institute for Science and Engineering, 
Waseda University, 3-4-1 Okubo, Shinjuku, Tokyo 169-8555, Japan}
\email{hiroki@heap.phys.waseda.ac.jp}

\begin{abstract}
 This paper is a sequel to our previous work
 for accretion onto a Schwarzschild black hole
 and the so-called standing accretion shock instability
 (SASI), in this paper we investigate non-axisymmetric
 perturbations for a Kerr black hole.
 The linear and non-linear phases for the shock evolution
 are analyzed in detail by both 2D general relativistic
 hydrodynamical simulations and linear analysis.
 Since the structure of steady axisymmetric accretion flows
 with a standing shock wave
 is very sensitive to the inner transonic flow,
 their properties such as Mach numbers,
 which are important for the stability,
 depend on the Kerr parameter very much.
 Although the essential features of the instability
 do not differ from the previous results
 for the Schwarzschild black hole,
 the frame dragging effects specific to
 the Kerr black hole is also evident.
 Interestingly, the oscillation periods
 of the fundamental unstable modes
 are dependent only on the shock radius
 irrespective of the injection parameters.

\end{abstract}


\section{Introduction}
 It is widely believed that
 the transonic rotating accretion flow onto a black hole
 is one of the candidate sources
 for many high energy astrophysical phenomena.
 The time variability of these objects may be interpreted as
 the unsteadiness of accretion flows,
 which in general suffer from various types of instabilities,
 such as thermal, viscous, dynamical, self-gravity,
 induced ones and so on.
 A variety of physical mechanisms (\citet{Varn2002,Varn2006,Don2008},
 and references theirin)
 have been considered to explain the observed intensity variations.
 The shock wave in the accretion flows
 is one of the promising mechanisms
 and the so-called ``shock oscillation model'' has been
 investigated by many authors
 \citep{Ch1989,molteni99,cha04,das03a,das03b,aoki04,molteni06,oku07}.
 The stability of accretion flows
 with a standing shock wave in them is one of the key issues.

 The linear stability of the standing shock wave
 against axisymmetric perturbations
 in the axisymmetric accretion flow
 is well-known and simply stated as follows:
 if the post-shock matter is accelerated,
 the shock is unstable \citep{nak94,nak95}.
 However, the accretion shock wave
 that is stable against axisymmetric perturbations
 can be unstable for non-axisymmetric perturbations
 \citep{molteni99,fog2003,Gu2005,nag08}.
 This non-radial shock instability may be quite common.
 In fact, the similar instability is found in accretion flows onto
 a nascent proto neutron star in the supernova explosion
 and might be an important element in the mechanism for it.

 Our previous paper (\citet{nag08} hereafter NYv1)
 investigated general relativistically
 the non-axisymmetric shock instability around a non-rotating black hole,
 suggesting also possible implications for fluctuations in GRB jets.
 However, there is no doubt
 realistic stellar mass black holes in general are rotating
 as suggested by the observations of massive stars \citep{fuku1982}.
 Since it is believed that rapidly rotating massive stars
 are the progenitor of long GRBs,
 the investigation of the standing accretion shock instability or SASI
 should be extended to the Kerr black hole.
 Moreover, in order to take into account the frame dragging
 due to the black hole's spin,
 the fully general relativistic treatment is strongly required.
 We observed in the previous paper that
 the general relativistic effects are important
 to determine the structure of the inner transonic flow.
 This in turn means that the injection parameters
 such as Bernoulli constant and specific angular momentum
 are also affected by the general relativistic effects.
 In fact, the injection parameters
 that allow the existence of the standing shock wave
 are sensitive to the Kerr parameters.

 The plan of this paper is as follows.
 In section 2, we describe the steady axisymmetric accretion flows
 with a standing shock wave in them.
 Then we present the formulation of
 linear analysis in section 3.
 The numerical methods for dynamical simulations and models
 are described in section 4.
 The results of the numerical simulations
 obtained by both linear analysis
 and dynamical simulations
 are shown in section 5.
 We conclude the paper in section 6 with summary and discussion.

\section{Axisymmetric steady accretion flows
 with a shock wave in Kerr spacetimes}
 The mathematical formalisms of axisymmetric steady accretion flows
 onto Kerr black hole with a standing shock wave
 are essentially the same as those used in NYv1.
 Only the equatorial plane is considered
 and the Kerr space metric is fixed in time.
 The conditions for the existence of
 multiple sonic points and standing shock waves in Kerr spacetimes
 were already discussed by \citep{lu986,lu997}.
 In the following we briefly review the essential points
 for constructing axisymmetric steady flows,
 which would help the reader to understand the property of
 the solutions.
 We refer to \citet{lu986,lu997} for more details.

 We first determine the range of the injection parameters
 such as Bernoulli constant and specific angular momentum,
 these give multiple sonic points.
 The existence of multiple sonic point is a necessary condition
 for the existence of standing shock wave,
 since not only the pre-shock flows but also post-shock flows are transonic.
 Then, two transonic accretion flows are connected by a standing shock wave,
 where the Rankine-Hugoniot relations are satisfied.
 Since the inner transonic flows are strongly
 affected general relativistic effects,
 the structure of the entire accretion flows
 also varies with Kerr parameters.

\begin{figure}
\epsscale{1.0}
\plotone{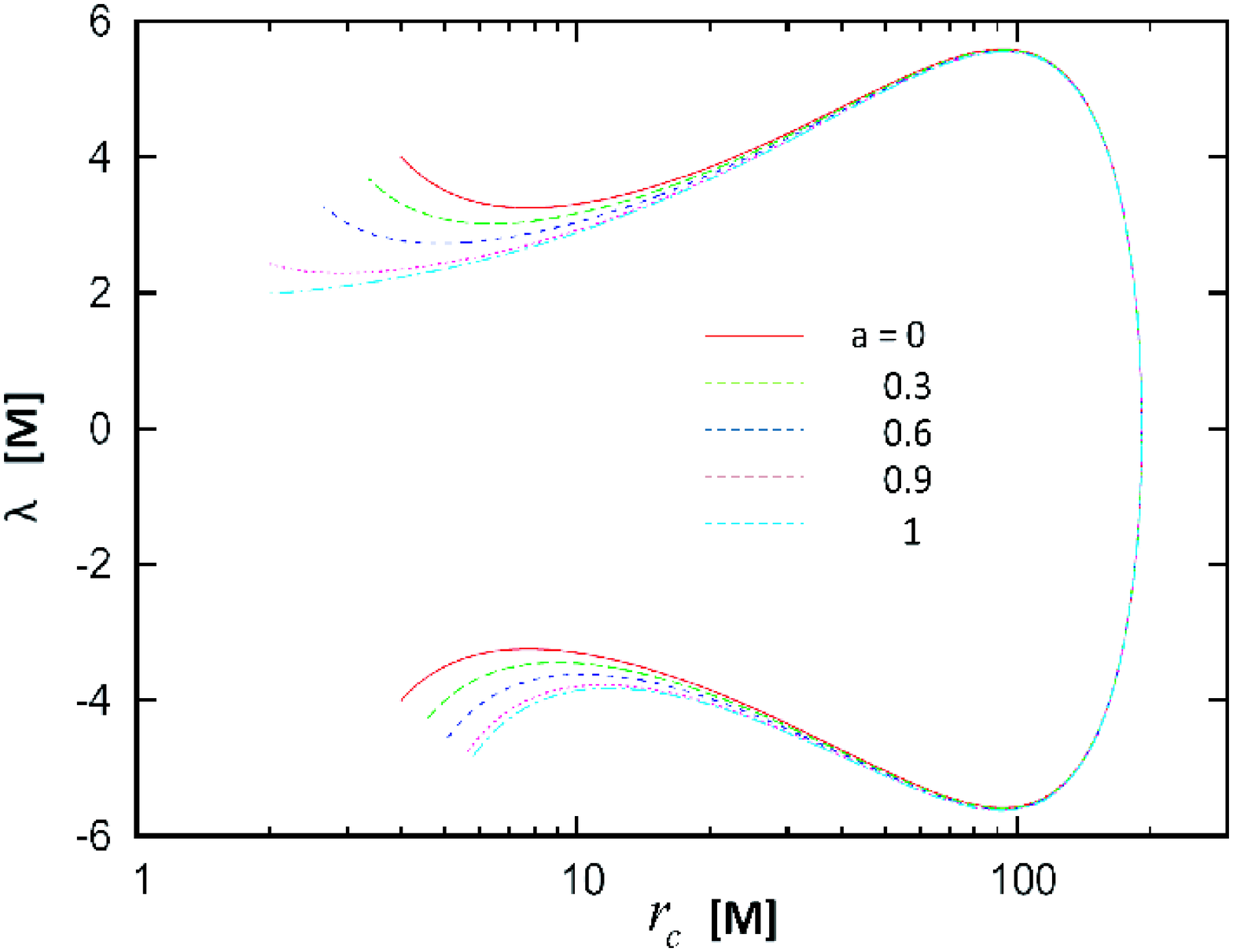}
\caption{The locations of sonic points as a function of
 the specific angular momentum and Kerr parameter.
 The adiabatic index and Bernoulli constant are fixed as
 $3/4$ and $1.004$, respectively.
 \label{ramdarc2note}}
\end{figure}

 The basic equations are the relativistic continuity equation and
 equation of energy-momentum conservations.
\begin{eqnarray}
\left(\rho_{0}u^{\mu}\right)_{;\mu} & = & 0 , \label{eq:continubase} \\
\left(T^{\mu\nu}\right)_{;\nu} & = & 0 . \label{eq:energymomentum}
\end{eqnarray}
 $\rho_{0},u^{\mu},T^{\mu \nu}$ are the rest mass density,
 4-velocity and energy momentum tensor, respectively.
 Throughout this paper,
 we assume a $\Gamma$-law~EOS,
 $p = \left(\Gamma-1\right)\rho_0 \epsilon$,
 where $p$ and $\epsilon$ are the pressure,
 and specific internal energy.
 Except for the discussion of dynamical simulations,
 we adopt the geometrical units $G=c=1$,
 where $G$ and $c$ are the gravitational constant
 and the speed of light.
 Latin and Greek indices denote spatial and spacetime components,
 respectively.
 Since we consider the accretion flows only in the equatorial plane,
 we ignore the $\theta$-component of velocity, $u^{\theta}$, and all $\theta$
 derivatives.
 The basic equations can be reduced to ordinally differential equations
 with respect to the radial coordinate:
\begin{eqnarray}
 \partial_{r} \left(r^2\rho_{0}u^r\right)
  & = & 0 , \label{eq:contibackground} \\
 \partial_{r} p + \rho_0u^{r}\partial_{r}\left(hu_r\right)
 & = & \frac{1}{2} \rho_0h \,
\biggl\{ \, \left( \partial_r g_{rr} \right) \left(u^r\right)^2 +
                \left( \partial_r g_{\phi\phi} \right) \left(u^{\phi}\right)^2 +
                \left( \partial_r g_{tt} \right) \left(u^t\right)^2 + \nonumber \\
   & &     \hspace{12mm} 2 \left( \partial_r g_{t{\phi}} \right) u^t u^{\phi} \, \biggr\} ,
 \label{eq:Eulerradialbackground} \\
 \partial_{r}\left(hu_t\right) & = & 0 , \label{eq:enmotimebackground} \\
 \partial_{r}\left(hu_{\phi}\right) & = & 0 , \label{eq:enmophibackground}
\end{eqnarray}
 where we use the Boyer-Lindquist coordinates.
 Here, $g_{\mu \nu}$, $h$ and $p$ are
 the metric, specific enthalpy and pressure, respectively.
 The only difference from the equations for the Schwarzschild black hole
 is the last term in eq.~(\ref{eq:Eulerradialbackground}),
 \{$2 \left( \partial_r g_{t{\phi}} \right) u^t u^{\phi}$\},
 which represents the frame dragging effect.
 It should be noted, however, that the frame dragging effect is included
 not only in this term but also in the 4-velocity itself.
 The transonic flow can be obtained by imposing
 the regularity conditions at the sonic point (see \citet{lu986}),
 where the equations become apparently singular.
 Figure~\ref{ramdarc2note} shows the locations of the sonic points
 as a function of the specific angular momentum and Kerr parameter.
 The adiabatic index and Bernoulli constant are fixed as
 $4/3$ and $1.004$, respectively in this figure.
 As is clearly demonstrated,
 the curve is symmetric with respect to $\lambda = 0$,
 which $\lambda$ denotes the specific angular momentum,
 for the Schwarzschild spacetimes,
 where as it is not true of the Kerr spacetimes.
 This is due to the frame dragging of Kerr black holes,
 which acts effectively as an additional angular momentum
 that forces a corotation of matter with the black hole,
 and thus the symmetry is broken.
 This figure also shows that
 there are indeed multiple sonic points
 for adequate injection parameters.
 It is well known that the physically-meaningful sonic points
 are the inner and outermost ones,
 which are so-called saddle-type singularities,
 while the unphysical middle sonic point is a central-type singularity.
 Hence, the steady accretion flows with a standing shock wave
 should pass through the inner- and outermost sonic points.
 The Rankine-Hugoniot relation determines the location of standing
 shock wave from these two transonic flows.
 These equations are completely the same as those used in the previous paper
 (see Section 2.2 in NYv1).
 Just like sonic points,
 the shock positions also vary with the Kerr parameter very much.

 As in the Schwarzschild black holes,
 there are generally two possible shock
 locations under adequate injection parameters.
 These are referred to as the inner and outer shock points.
 We consider only the outer shock in this paper,
 since it is well known that the inner shock
 is already unstable against axisymmetric perturbations,
 which is also true of the Kerr spacetimes \citep{nak95}.
 We construct axisymmetric steady accretion flows imbedding on
 outer shock wave for different injection and Kerr parameters,
 which are summarized in Table~\ref{initabnote}.
 The adiabatic index is fixed to $\Gamma = 4/3$
 throughout the paper.
 It should be noted that
 some models have negative specific angular momentum,
 i.e, matter rotates in the opposite direction of the black hole spin,
 although it is rather unlikely in realistic astrophysical systems.
 It is, however, useful to elucidate the Kerr parameter dependence
 and the following fact, which will be demonstrated shortly,
 also gives another merit:
 as the Kerr parameter gets larger,
 the range of the injection parameters
 that allows the existence of a standing shock wave
 become smaller and disappears eventually in the corotation case,
 whereas it is not true for the counter rotation case.
 This allows us to investigate the properties of the instability
 under wider injection parameters.

\begin{figure}
\epsscale{1.30}
\plotone{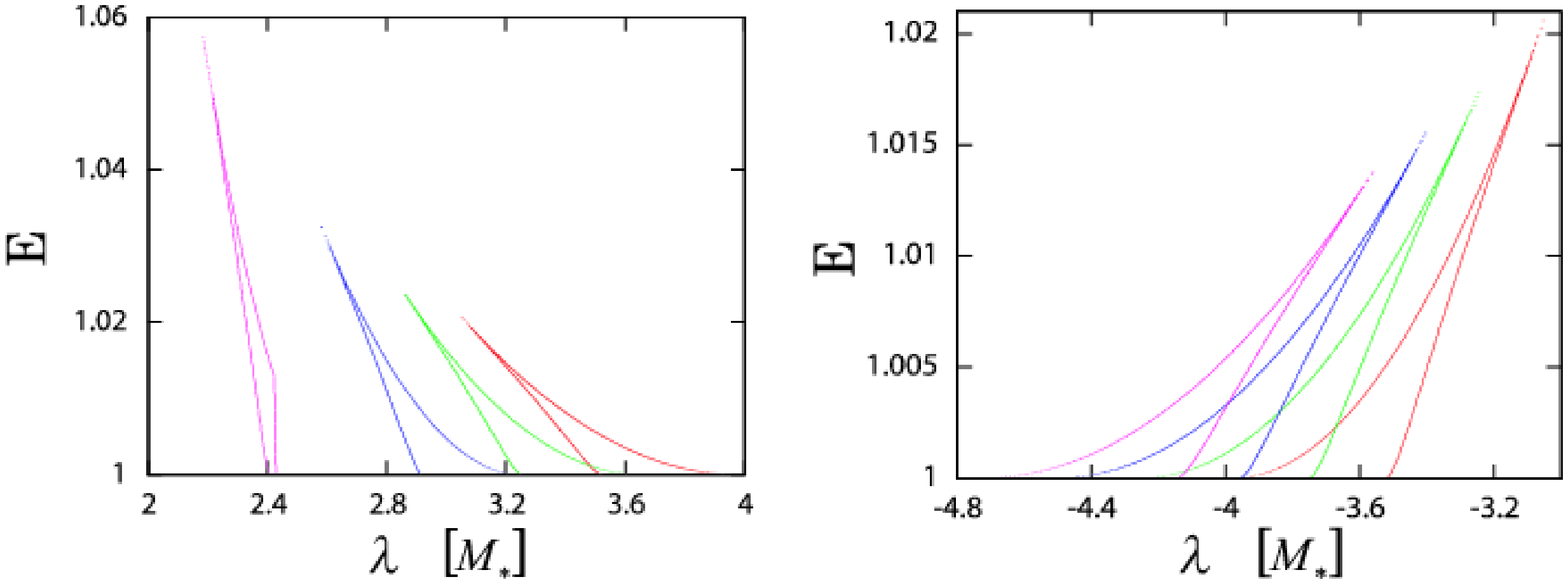}
\caption{The ranges of the injection parameters that allow the existence
 of a standing accretion shock wave around Kerr black holes.
 The red, green, blue and pink lines show the boundaries
 of the allowed regions for
 $a/M_{\ast}=0$, $0.3$, $0.6$ and $0.9$, respectively.
 The left panel shows the corotation case and
 the right panel gives the counter-rotation case.
 \label{combineshockrangenote}}

\end{figure}

 In Figure~\ref{combineshockrangenote},
 we show the ranges of injection parameters,
 for which the standing shock wave can exist.
 As seen in the left panel, where the corotation case is presented,
 as the Kerr parameter becomes larger,
 the allowed regime of the specific angular momentum
 shifts to smaller values.
 This is due to the frame-dragging by the Kerr black hole,
 which acts effectively to increase the angular momentum of matter.
 Note that the specific angular momentum is conserved along the stream line
 and frame-dragging effect is not restricted to the vicinity of black hole.
 The boundary of the allowed regions normally consists of two areas.
 For a given Bernoulli constant,
 the maximum angular momentum is determined by the weak shock limit,
 i.e, Mach number of the shock $M_{sh} \sim 1$.
 On the other hand, the minimum value is determined by the location
 of the outer shock wave.
 It is widely known that the outer shock wave is located at a larger radius
 than the middle sonic point,
 while the inner shock wave is located at a smaller radius.
 If the specific angular momentum becomes smaller,
 the locations of each shock wave approach the middle sonic point,
 and eventually coincide at this point.
 This determines the minimum specific angular momentum for
 the existence of a standing shock wave.
 As mentioned in NYv1, however,
 for small Bernoulli constants,
 there is a maximum specific angular momentum
 above which no multiple sonic points exist.
 This is true only the fully general relativistic treatment.
 Hence, there is a possibility that
 the maximum specific angular momentum is determined
 by this limit.
 As a matter of fact, there is indeed the case of $a/M_{\ast}=0.9$,
 in which the allowed regime is bounded
 by a vertical like for small Bernoulli constants.

 The right Figure~\ref{combineshockrangenote} shows the
 region of injection parameters, for which the steady shock wave exists
 in the conter-rotation case,
 where matter rotates in the opposite direction of the Kerr black hole spin.
 As expected, the allowed regions are quite different from
 those for the corotation case.
 As the Kerr parameter becomes large,
 the specific angular momentum gets more negative
 and the Bernoulli constant becomes smaller.
 Because the frame dragging effects decelerate the matter rotation,
 the specific angular momentum has to be more negative.
 As mentioned above, the specific angular momentum
 is conserved along the stream line,
 it has too large an absolute value
 (and hence too great centrifugal force)
 at distant places from the black hole,
 where the frame dragging is negligible.
 The excess should be compensated for
 by reducing the Bernoulli constant
 and as a result, pressure.

\section{Linearized equations in Kerr spacetimes}
 We perform a linear analysis for the non-axisymmetric
 shock instability around a Kerr black hole.
 The obtained eigen-values help us to analyze
 not only the shock stability,
 but also validate the accuracy of dynamical simulations.
 By combining dynamical simulations with linear analysis,
 both the linear and non-linear phases can be investigated in detail.
 In this section, we present a formulation
 for the linear analysis including the boundary conditions.

 The basic equations are the linearized relativistic continuity
 and energy-momentum conservation equations.
 The perturbed quantities are assumed to be proportional to
 $e^{-i\omega t + im\phi}$.
 Since we consider the equatorial plane only
 and neglect all the $\theta$-derivatives and
 the $\theta$-component of velocity,
 the linearized equations are expressed as follows:
\begin{eqnarray}
\partial_rf & = & 
\frac{i}{\rho_{0(0)} u^{r(0)}}\biggl\{ \,\rho_{0(1)}u^{t(0)}\sigma +
                       \rho_{0(0)}\left( \omega u^{t(1)}-mu^{\phi(1)} \right)\, \biggr\} ,
 \label{eq:lineconti2} \\
\partial_rq & = & \frac{i}{\rho_{0(0)}u^{r(0)}h_{(0)}u^{t(0)}}
 \left(\, \omega p_{(1)} + \rho_{0(0)} u^{t(0)} h_{(0)} u_{t(0)}
 \sigma q \,\right) , \label{eq:ber2}\\
\partial_rV_{(1)} & = & 
i \frac{u^{t(0)}}{u^{r(0)}} \sigma V_{(1)} , \label{eq:linework2} \\
\partial_rS_{(1)} & = & 
i \frac{u^{t(0)}}{u^{r(0)}} \sigma S_{(1)} , \label{eq:lineentro2}
\end{eqnarray}
where $S$ is the entropy per baryon and the following notations are used:
\begin{eqnarray}
f & \equiv & \frac{\rho_{0(1)}}{\rho_{0(0)}} + \frac{u^{r(1)}}{u^{r(0)}} , \label{eq:fdef} \\
q & \equiv & \frac{h_{(1)}}{h_{(0)}} + \frac{u_{t(1)}}{u_{t(0)}} , \label{eq:qdef} \\
V_{(1)} & \equiv & \omega\left(\, h_{(1)} u_{\phi(0)} + h_{(0)}u_{\phi(1)} \,\right)
+ m\left(\, h_{(1)} u_{t(0)} + h_{(0)} u_{t(1)} \,\right) , \label{eq:Vdef} \\
\sigma & \equiv & \omega - m \frac{u^{\phi(0)}}{u^{t(0)}} . \label{eq:sigmadef}
\end{eqnarray}
 $f$ and $q$ are the values which relate to the mass flux and Bernoulli constant,
 respectively.
 $V_{1}$ is a useful value which we can integrate analytically.
 The $\sigma$ denotes the complex frequency of the perturbations
 measured in the rotating frame.
 Note that these equations for the Kerr spacetimes
 look completely the same as those used in NYv1
 for the Schwarzschild spacetimes.
 As mentioned in the previous section, however,
 the frame-dragging effects are included implicitly
 in the components of 4-velocity.
 As in the Schwarzschild black hole case,
 $V_{(1)}$ and $S_{(1)}$ can be integrated analytically
 (see Section 3 in NYv1), and we only need to integrate numerically
 Eqs.(\ref{eq:lineconti2}) and (\ref{eq:ber2}).

 The boundary conditions are also the same as those used in NYv1.
 We impose the outer boundary condition on
 the linearized Rankine-Hugoniot relation at the shock surface,
 assuming that the pre-shock region remains unperturbed.
 The explicit forms of these equations are given by,
\begin{eqnarray}
 \left( \rho_{0(0)} u^{r(1)} \right)_{+} + \left( \rho_{0(1)} u^{r(0)} \right)_{+} + A = 0 ,
 \label{eq:lineshocksurface1} \\
 \biggl\{ \left(\rho_0 h u_t u^r\right)^{(0)} \, 
    \left( \frac{\rho_{0(1)}}{\rho_{0(0)}} 
    + \frac{h_{(1)}}{h_{(0)}}
    + \frac{u_{t(1)}}{u_{t(0)}}
    + \frac{u^{r(1)}}{u^{r(0)}}       \right) \biggr\}_{+} + B = 0 ,
 \label{eq:lineshocksurface2} \\
\biggl\{  \left( \rho_0 h u_r u^r \right)^{(0)} \,
    \left( \frac{\rho_{0(1)}}{\rho_{0(0)}} 
    + \frac{h_{(1)}}{h_{(0)}}
    + \frac{u_{r(1)}}{u_{r(0)}}
    + \frac{u^{r(1)}}{u^{r(0)}} \right) + p_{(1)} \biggr\}_{+} + C = 0 ,
 \label{eq:lineshocksurface3} \\
 \biggl\{ \left(\rho_0 h u_{\phi} u^r\right)^{(0)} \, 
    \left( \frac{\rho_{0(1)}}{\rho_{0(0)}} 
    + \frac{h_{(1)}}{h_{(0)}}
    + \frac{u_{\phi(1)}}{u_{\phi(0)}}
    + \frac{u^{r(1)}}{u^{r(0)}}       \right) \biggr\}_{+} + D = 0 ,
 \label{eq:lineshocksurface4}
\end{eqnarray}
with the following definitions:
\begin{eqnarray}
 A &\equiv&  i \omega \eta \,
          \biggl\{ \left( \rho_{0(0)} u^{t(0)} \right)_{+} -
             \left( \rho_{0(0)} u^{t(0)} \right)_{-}  \biggr\}
           - i m \eta \,
          \biggl\{ \left( \rho_{0(0)} u^{\phi(0)} \right)_{+} -
             \left( \rho_{0(0)} u^{\phi(0)} \right)_{-}  \biggr\} ,
          \label{defworkA} \\
 B &\equiv&  i \omega \eta \,
          \biggl\{  \left(\rho_{0} h u_{t} u^{t} + p \right)^{(0)}_{\!+} -
              \left(\rho_{0} h u_{t} u^{t} + p \right)^{(0)}_{\!-}  \biggr\} \nonumber \\
&&           - i m \eta \,
          \biggl\{  \left(\rho_{0} h u_t u^{\phi}\right)^{(0)}_{\!+} -
              \left(\rho_{0} h u_t u^{\phi}\right)^{(0)}_{\!-}  \biggr\} ,
          \label{defworkB} \\
 C &\equiv& \biggl\{ \left( \frac{d}{dr}
               \left( \rho_{0(0)} h_{(0)} u_{r(0)} u^{r(0)}
                      + p^{(0)}  \right)\right)_{+}
              -\left( \frac{d}{dr}
               \left( \rho_{0(0)} h_{(0)} u_{r(0)} u^{r(0)}
                      + p^{(0)}  \right)\right)_{-} \biggr\}
             \, \eta
 \nonumber \\
&&          + i \omega \eta \,
          \biggl\{  \left(\rho_{0} h u^t u_r \right)^{(0)}_{\!+} -
              \left(\rho_{0} h u^t u_r \right)^{(0)}_{\!-}  \biggr\} 
            - i m \eta \,
          \biggl\{  \left(\rho_{0} h u_r u^{\phi}\right)^{(0)}_{\!+} -
              \left(\rho_{0} h u_r u^{\phi}\right)^{(0)}_{\!-}  \biggr\} ,
          \label{defworkC} \\
 D &\equiv&  i \omega \eta \,
          \biggl\{  \left(\rho_{0} h u_{\phi} u^t \right)^{(0)}_{\!+} -
              \left(\rho_{0} h u_{\phi} u^t \right)^{(0)}_{\!-}  \biggr\} 
 \nonumber \\
&&          - i m \eta \,
          \biggl\{  \left(\rho_{0} h u_{\phi} u^{\phi} + p \right)^{(0)}_{\!+} -
                    \left(\rho_{0} h u_{\phi} u^{\phi} + p \right)^{(0)}_{\!-}  \biggr\} .
          \label{defworkD}
\end{eqnarray}
 where $\eta$ denotes the displacement of the shock radius and defined as:
\begin{eqnarray}
 R_{sh} = R_{sh(0)} +
 \eta \exp\left(\,-i \omega t + i m \phi \,\right) ,
 \label{eq:Rshpertu}
\end{eqnarray}
 where $R_{sh}$ stands for the shock radius.

 Finally, the regularity condition is imposed at the inner sonic point
 as the inner boundary condition.
 The explicit form of this condition is;
\begin{eqnarray}
 G_{(1)} - F_{(1)}u_{r(0),r} = 0 , \label{eq:regso}
\end{eqnarray}
 where the following notations are employed
\begin{eqnarray}
F_{(0)} & \equiv & \rho_{0(0)} h_{(0)} g^{rr}
\biggl\{  u^{r(0)} u_{r(0)} - \left(b_{s(0)} \right)^2 \left( 1 + u^{r(0)} u_{r(0)} \right)  \biggr\}
 \label{eq:Fbackground} \\
F_{(1)} & \equiv & \, 
   \rho_{0(1)} h_{(0)} \left(u^{r(0)}\right)^2
 + \rho_{0(0)} h_{(1)} \left(u^{r(0)}\right)^2
 + 2 \rho_{0(0)} h_{(0)} u^{r(0)} u^{r(1)} \nonumber \\
&&  \, 
 - \Gamma g^{rr} \biggl\{ p_{(1)} \left( 1+u^{r(0)} u_{r(0)} \right)
                   + 2 p_{(0)} u^{r(1)} u_{r(0)} \biggr\} ,
   \label{eq:woregso2} \\
 G_{(1)} & \equiv & \, \Gamma \left( g^{rr} \! _{,r} + \frac{2}{r} \right)
 \biggl\{ \left( p_{(1)} u_{r(0)} + p_{(0)} u_{r(1)} \right)\left( 1 + u_{r(0)} u^{r(0)} \right)
   + 2 p_{(0)} \left( u_{r(0)} \right)^2 u^{r(1)}  \biggr\} \nonumber \\
 &&  + \, \frac{1}{2} \biggl\{ \rho_{0(1)} h_{(0)} u^{r(0)} 
                   + \rho_{0(0)} h_{(1)} u^{r(0)}
                   + \rho_{0(0)} h_{(0)} u^{r(1)} \biggr\}  \nonumber \\
 &&    \biggl\{  g_{rr,r} \left(u^{r(0)}\right)^2 
      + g_{\phi \phi ,r} \left(u^{\phi(0)}\right)^2
      + g_{ t t ,r } \left(u^{t(0)}\right)^2
      + 2 g_{t \phi ,r} u^{t(0)} u^{\phi(0)}  \biggr\}  \nonumber \\
 &&  + \,  \rho_{0(0)} h_{(0)} u^{r(0)}
      \biggl\{ g_{rr,r} u^{r(0)} u^{r(1)} +
         g_{\phi \phi ,r} u^{\phi(0)} u^{\phi(1)} +
         g_{tt,r} u^{t(0)} u^{t(1)} +                  \nonumber \\
 &&      g_{t \phi ,r} \left( u^{t(0)} u^{\phi(1)} + u^{t(1)} u^{\phi(0)}\right)
                                                                 \biggr\}
     + \, i \rho_{0(0)} h_{(0)} u^{r(0)} u^{t(0)} u_{r(1)} \sigma \nonumber \\
 &&  + \,  i \Gamma p_{(0)} \left( 1 + u_{r(0)} u^{r(0)} \right)
                       \left( - \omega u^{t(1)} + m u^{\phi (1)} \right) \nonumber \\
 &&  -\,  i u^{t(0)} p_{(1)} \sigma  . \label{eq:woregso1}
\end{eqnarray}
 The unperturbed sound velocity in the comoving frame and
 the adiabatic index are denoted by $b_{s(0)}$ and $\Gamma$, respectively.
 Note again that the only explicit modifications
 from the Schwarzschild spacetime case are
 the terms with $g_{t \phi}$ (see section 3 in NYv1),
 although the frame dragging effects are included implicitly in the 4-velocity.

\section{Numerical Methods and Models}
 We perform two dimensional general relativistic hydrodynamical
 simulations of non-axisymmetric shock instability
 in the equatorial plane around a Kerr black hole.
 The numerical methods in the present paper are essentially the same
 as those used in NYv1.

 Our numerical code is based on the so-called central scheme,
 which guarantees a good accuracy even if flows include shock waves.
 We refer the reader to the Appendix of NYv1 for more details
 on this code.
 The Kerr-Schield coordinates, which have no coordinate singularity
 and allow us to put the inner boundary inside the event horizon,
 are used.
 The computational domain of $ 0.6 r_{eve}  \leq r \leq 200M_{\ast}$
 is covered by the $600(r) \times 60(\phi)$ grid points,
 where $M_{\ast}$ and $r_{eve}$ stand for the black hole mass
 and the radius of the event horizon in
 the equatorial plane, respectively.
 The radial grid width is non-uniform,
 increasing by $0.34\%$ per zone
 from the smallest grid ($ \Delta r = 0.1 M_{\ast}$)
 at the inner boundary.
 The computational times are $ \sim 6 \times 10^4 M_{\ast}$,
 which corresponds to $\sim 500$ms for $M_{\ast}=3$M$_{\sun}$.
 In the following, we discuss the numerical results
 in cgs units for the case $M_{\ast}=3$M$_{\sun}$,
 since we have in mind the application to massive stellar collapses
 as in NYv1.
 Note, however, that the formulations are dimensionless,
 and the scaling is obvious.

 The initial perturbation mode has $m=1$, where $m$ indicates
 the azimuthal mode number in $e^{im\phi}$,
 and the amplitude is fixed to $1\%$
 (see NYv1 for the dependence on the mode number and the initial amplitude).
 The radial distributions of initial perturbations are
 obtained by the linear analysis.
 We choose the most unstable $m=1$ mode,
 which allows us to compare clearly the linear growth phase
 between the linear analysis and dynamical simulations.

\section{Numerical Results}
 In this section, we first describe
 the results of the linear analysis,
 which are obtained by solving Eqs.
 (\ref{eq:lineconti2})-(\ref{eq:lineentro2})
 with the appropriate boundary conditions
 at the shock surface and sonic point
 (see Section~3).
 Then, we present the time evolutions of shock instability obtained by
 GRHD simulations in both linear and non-linear phases
 compared them with the linear analysis.
 In so doing,
 we employ the mode decomposition on
 the shock surface by the Fourier transform:
\begin{equation}
 a_m\left(t\right) = \int_0^{2\pi} R_{sh}\left(\phi,t\right) \,
 e^{im\phi} \, d\phi ,
\end{equation}
 where $R_{sh}\left(\phi,t\right)$ and $a_m\left(t\right)$ are
 the radius of the shock wave as a function of $\phi$ and $t$
 and the amplitude of mode $m$ as a function of $t$, respectively.

\subsection{Linear Analysis}
\begin{figure}
\epsscale{1.50}
\plotone{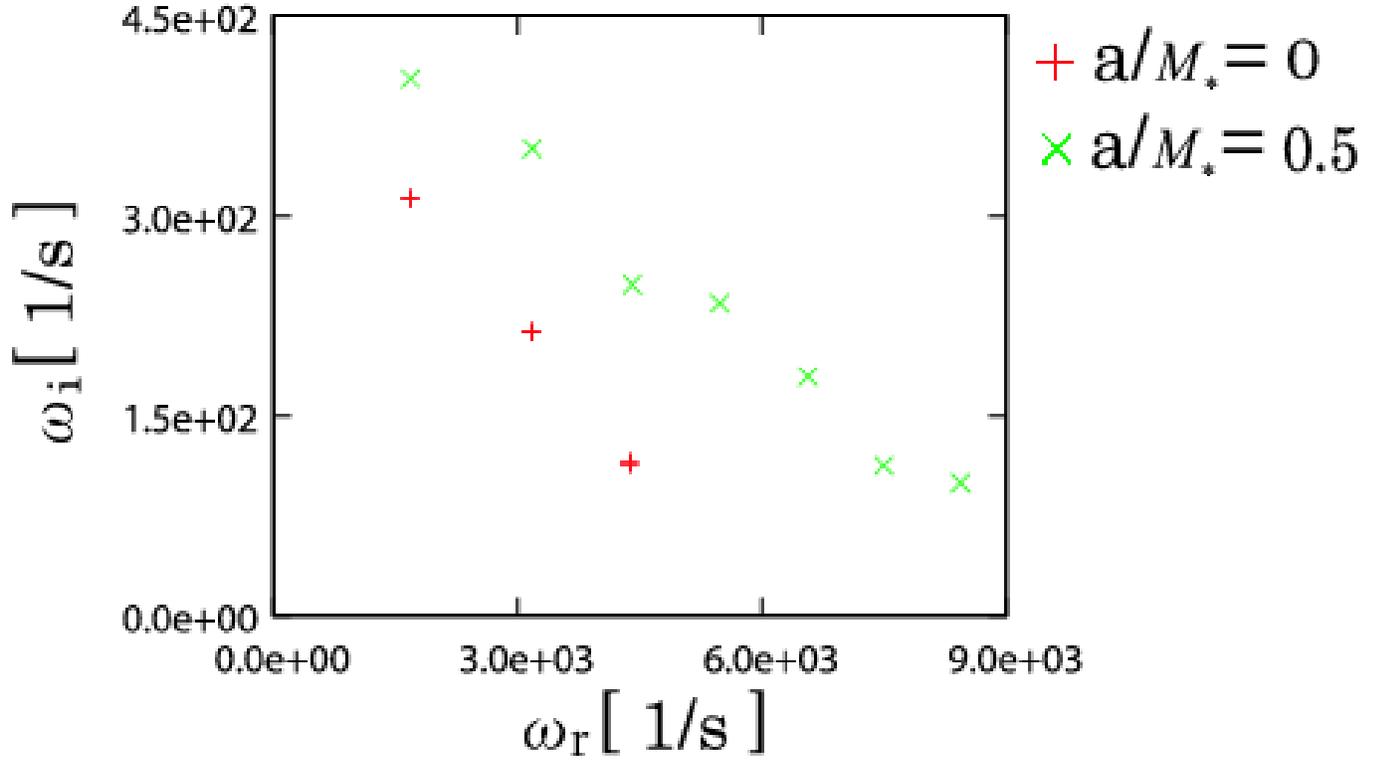}
\caption{The eigen-frequencies of unstable $m=1$ mode
  for models Mk05E4L301p (standard model) and Mk00E4L343p.
  These models have the same shock location ($r=16M_{\ast}$)
  and Bernoulli constant ($E=1.004$),
  but a different Kerr parameter and specific angular momentum.
 \label{graphomegastandardmodnote}}
\end{figure}

 The real and imaginary parts of unstable eigen-frequencies
 for the standard Model Mk05E4L301p
 are given in Figure~\ref{graphomegastandardmodnote},
 which also shows the another parts for Model Mk00E4L343p for comparison.
 In these models, the Bernoulli constant is identical and
 the Kerr parameter and specific angular momentum are chosen
 so that the shock radius would become the same.
 As seen in Figure~\ref{graphomegastandardmodnote},
 there are the unstable eigen-frequencies for Model Mk05E4L301p,
 and we also find that the higher frequency is stable than the lower frequency
 among these models.
 It should be noted, however, that this property is not always satisfied
 (see Fig.5 in NYv1).
 The magnitude of the imaginary part of omega
 is slightly larger than the Mk00E4L343p case,
 because the shock wave for Mk05E4L301p is stronger than Mk00E4L343 case.

\begin{figure}
\epsscale{1.5}
\plotone{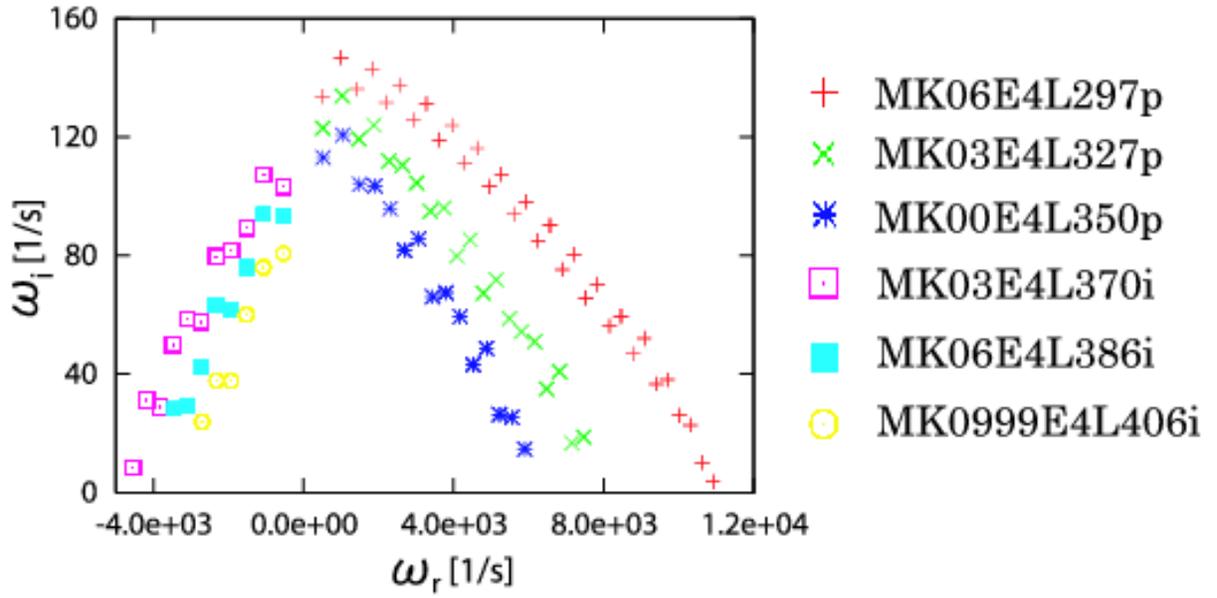}
\caption{The same as Figure~\ref{graphomegastandardmodnote},
 but for different models.
 The shock location of these models is fixed to $r=35M_{\ast}$.
 Note that the negative values of $\omega_{r}$ correspond to
 the conter-rotating models.
 \label{instaomefixshocknote}}
\end{figure}

 Figure~\ref{graphomegastandardmodnote} also tells us
 the dependence on the Kerr parameter:
 the flow tends to be more unstable
 as the Kerr parameter gets larger.
 Note that the Bernoulli constant
 is not changed and the specific angular momentum is
 chosen in such a way that
 the shock location is unchanged.
 This trend is also seen in other models.
 In Figure~\ref{instaomefixshocknote}, we show that
 the real and imaginary parts of unstable eigen-frequencies
 for models Mk00E4L350p to Mk0999E4L406i (See Table~\ref{initabnote}),
 have the same shock position ($R_{sh}=35M_{\ast}$).
 As can be clearly seen in this figure,
 the imaginary part of the eigen-frequency
 becomes larger in the corotation case,
 while it becomes smaller in the inverse rotation case
 (with increasing Kerr parameters).
 The reason is as follows.

 From Table~\ref{initabnote},
 we find that the growth rates and the Mach numbers
 of the pre-shock flows are correlated with each other.
 Stronger shocks tend to have a larger imaginary part of eigen-frequency,
 i.e., they are more unstable.
 This fact is the key to the above trend.
 The inner transonic flow is strongly affected by the frame dragging
 of the Kerr black hole and it works effectively
 as an addition to the specific angular momentum of matter.
 As a result, the specific angular momentum becomes smaller
 with increasing Kerr parameters if the shock location is fixed.
 Since the outer transonic flow is hardly affected
 by the frame dragging effect,
 this leads to the reduction of the centrifugal force there.
 As a result, the radial velocity of outer transonic flow becomes
 faster (slower) in the corotation (counter-rotation) case,
 and the Mach number of the pre-shock flow gets larger (smaller)
 with increasing Kerr parameters,
 which then leads to the trend that
 the models with larger Kerr parameters 
 tend to be more unstable (stable)
 in the corotation (counter-rotation) case.

\begin{figure}
\epsscale{.80}
\plotone{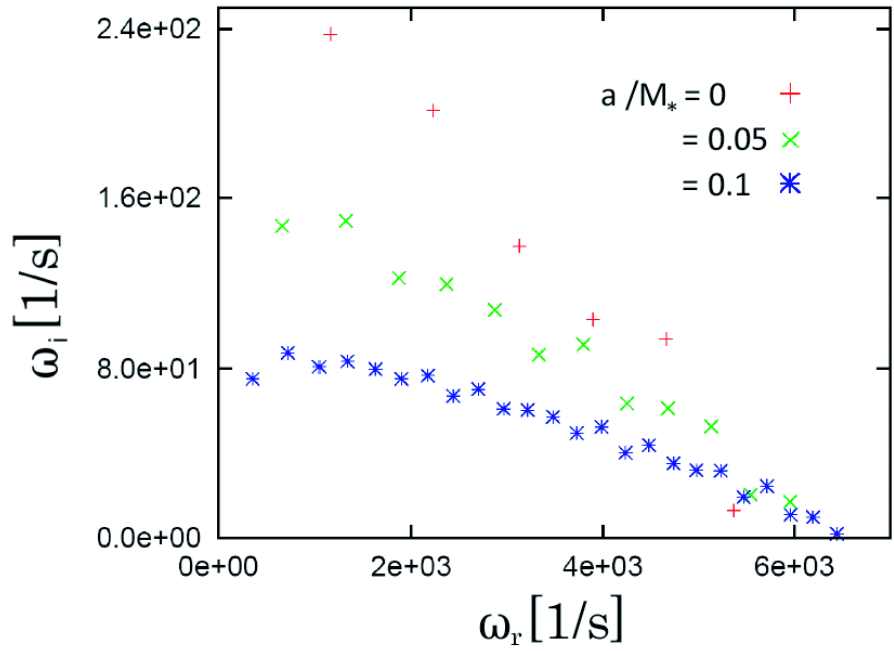}
\caption{The eigen-frequencies of unstable models for the models
 with the same injection parameters,
 but different Kerr parameters.
 \label{fixinjecdiffeKerrnote}}
\end{figure}

 It is also interesting to present the dependence of shock stability
 on the Kerr parameter under
 {\it the same injection parameters}.
 In Figure~\ref{fixinjecdiffeKerrnote},
 we shows the eigen-frequencies for models Mk00E4L345p,
 Mk005E4L345p and Mk01E4L345p, all corotation cases.
 The initial steady flows for these models are constructed
 with the same injection parameters but different Kerr parameters.
 As found from Table~\ref{initabnote},
 the shock radius becomes larger
 with increasing Kerr parameters in this case.
 Recalling our previous results
 for the Schwarzschild black hole,
 that the shock radius becomes larger
 as the specific angular momentum is
 increased with the same Bernoulli constant,
 we can interpret the present results as follows:
 as the Kerr parameter increases,
 the specific angular momentum becomes effectively larger,
 and the shock is pushed outwards.
 Then the Mach number of pre-shock flows becomes smaller,
 since the radial velocity is a decreasing function of radius.
 As a result, the flow becomes more stable,
 or $\omega_{i}$ gets smaller.

 In Figures~\ref{instaomefixshocknote} and \ref{fixinjecdiffeKerrnote},
 we find that the unstable eigen-frequencies for higher harmonics 
 oscillate for some models.
 We think that this is not a numerical artifact,
 and the oscillatory behavior is found in preceding
 papers (NYv1, \citet{fog2003}).
 At present, the reason for these oscillations is unclear,
 and further investigations are necessary.

\begin{figure}
\epsscale{1.0}
\plotone{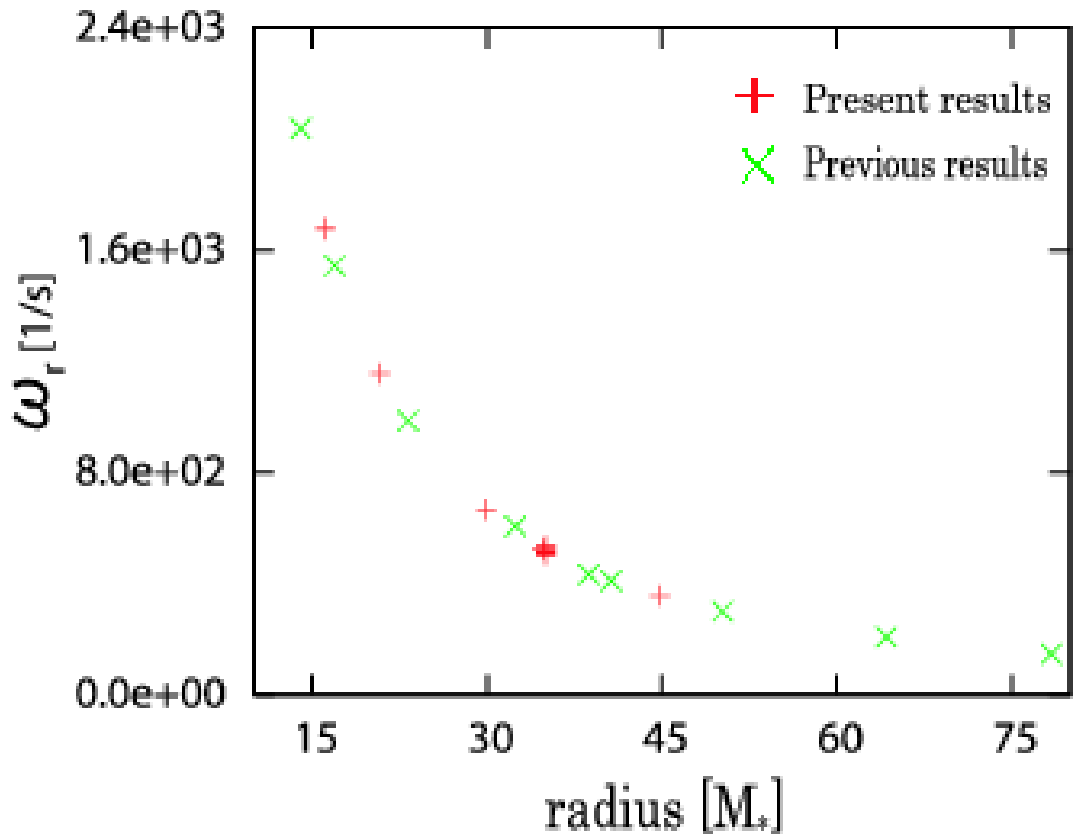}
\caption{The shock radii and the real parts of the eigen-frequencies
 for the fundamental unstable modes for various models.
 In addition to the present results (green dots),
 the previous results (NYv1) are included as red dots.
 \label{graphomeshockrelationnote}}
\end{figure}

 Finally, we discuss a possible strong correlation
 between $\omega_{r}$ for the fundamental unstable modes
 and the shock radius.
 In Figure~\ref{graphomeshockrelationnote}, we show
 the shock radii and $\omega_{r}$ for the fundamental unstable
 modes for all the various models.
 We also include the previous results (NYv1) in this figure
 to increase the number of samples.
 The injection parameters for the previous results
 are listed in Table~1 in NYv1.
 As is evident, the real part of fundamental unstable modes
 decreases monotonically with increasing shock radius
 even though all models does not have the same injection parameters,
 adiabatic indices or Kerr parameters.
 This suggests that the shock radius is the unique factor to decide
 the oscillation frequency.
 It also means that
 the rotational pattern velocity of the deformed shock surface,
 which corresponds to $\omega_{r}$,
 is a function of the shock radii alone.
 It is interesting that
 the specific angular momentum has no direct effect on it.
 The relation will also be useful to infer $\omega_{r}$
 from the shock radius without detailed calculations.
 At present, we do not understand the reason for this correlation.
 This may be a clue to the understanding of the mechanism SASI,
 and definitely needs further investigations.

\subsection{Numerical Simulations}
\begin{figure}
\epsscale{.80}
\plotone{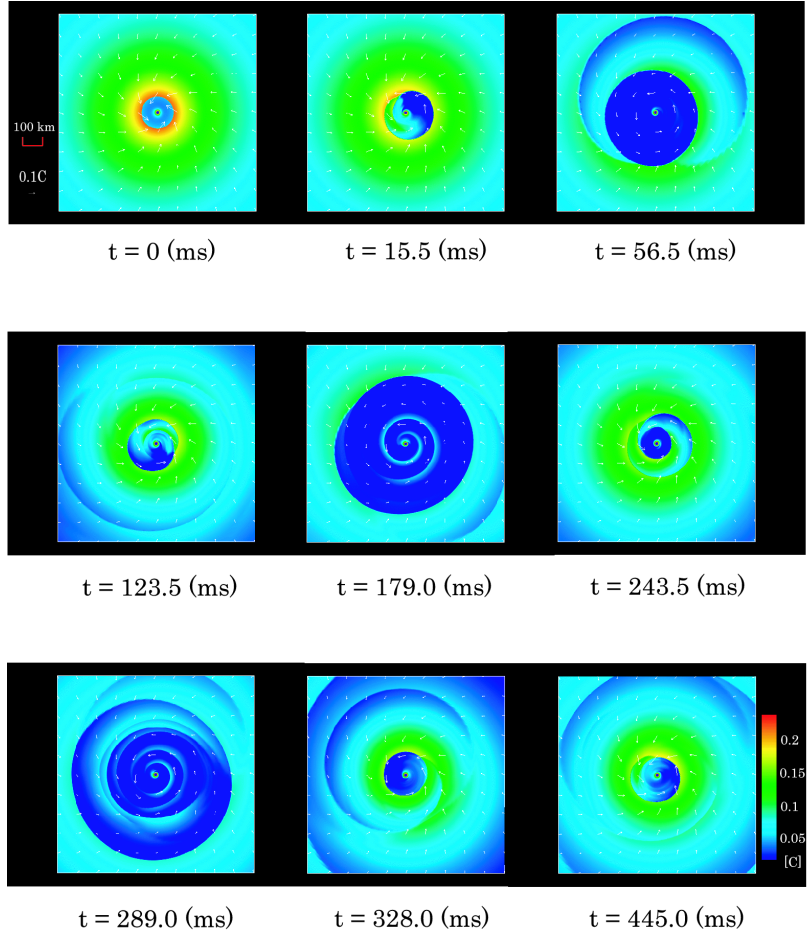}
\caption{The time evolution of the velocity for Model Mk05E4L301p.
 The color contour shows the magnitude of radial velocity.
 The arrows represent the velocities at their positions.
 The central region in blue corresponds to the black hole.
 \label{Kerrfigforpapver3note}}
\end{figure}

 Although we investigate several models with different Kerr parameters,
 the qualitative evolutionary path
 is generally similar to that found in the previous simulations NYv1.
 We first summarize the basic features in the temporal evolution
 of the standard Model Mk05E4L301p,
 which has the same Bernoulli constant and shock location
 but a Kerr parameter and specific angular momentum different
 from Model Mk00E4L343p,
 which is the standard model M1 in the previous paper (NYv1).
 Figure~\ref{Kerrfigforpapver3note} shows
 the snapshots of the velocity for this model.
 In the linear phase,
 a purely single $m=1$ mode grows exponentially
 and the deformed shock wave starts to form a spiral arm, which
 rotates in the same direction as the unperturbed flow.
 Over times, the spiral arm develops and
 non-linear mode couplings can not be negligible.
 When the non-linear regime is reached,
 the dynamics becomes highly complex.
 Several shocks start to collide with each other and
 the original shock begins to oscillate radially.
 While large and small radial oscillations
 are repeated like a limit cycle,
 we find that quasi-steady of asymmetric configurations $m=1$ or $2$
 are realized eventually for most of the models.
 For some models that have too large or too small radial oscillations,
 such as models Mk01E1L350p and Mk02E8L350i,
 we can not identify the dominant mode in the non-linear phase.

\begin{figure}
\epsscale{.80}
\plotone{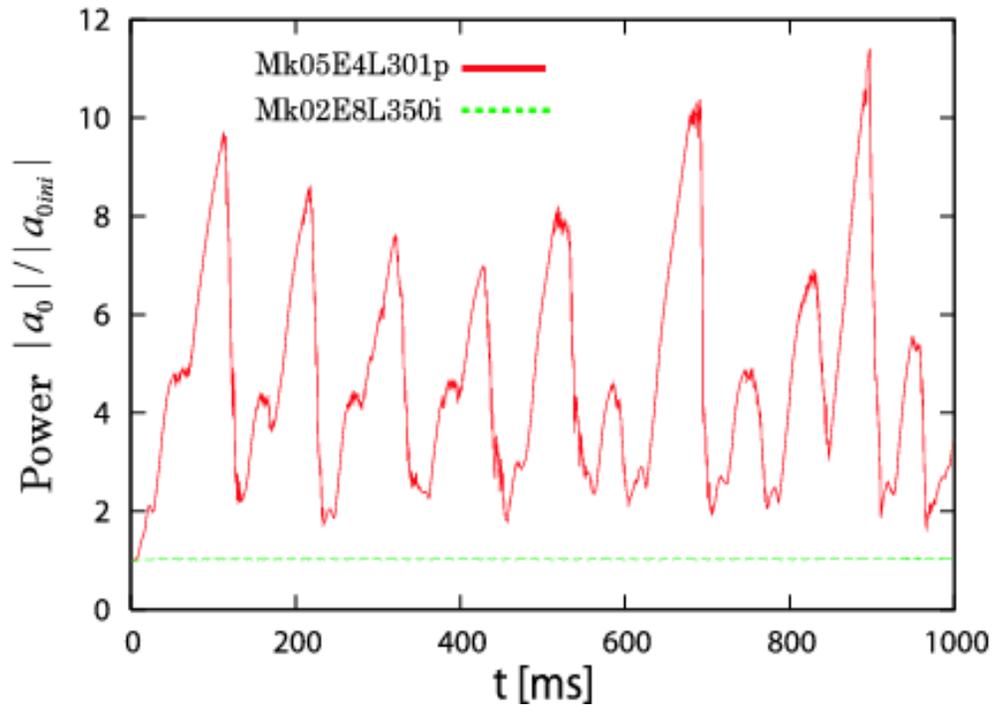}
\caption{ The evolutions of the amplitude of $m=0$ mode for Mk02E8L350i
 which we evolve long time (1000ms) simulations.
 The same evolutions for standard Model Mk05E4L301p
 are also shown for comparison.
 \label{graphevom0_Mk02E8}}
\end{figure}

 In Table~\ref{instapropenote},
 we give the saturation amplitudes of $m=0$ modes,
 which vary from model to model.
 We recognize a strong correlation
 between the Mach number and the saturation amplitude
 of axisymmetric $m=0$ mode:
 stronger shocks tend to give larger saturation amplitudes.
 As a matter of fact, the shock wave for Model Mk01E1L350p,
 which was the largest Mach number,
 leaves the computational domain soon
 after the beginning of numerical simulations.
 On the contrary, the shock wave for Model Mk02E8L350i
 with the smallest Mach number does not move so much
 from the initial position throughout the simulation
 (see Figure~\ref{graphevom0_Mk02E8}).

\begin{figure}
\epsscale{.80}
\plotone{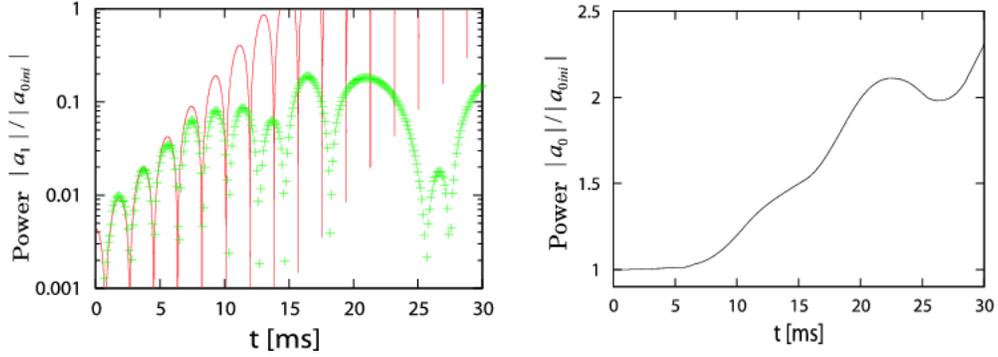}
\caption{The time evolution for Model Mk05E4L301p.
 Left: the comparison of the time evolutions
 of the amplitudes of the $m=1$ mode
 between the linear analysis (red lines)
 and dynamical systems (green crosses).
 Right: the evolutions of the amplitude of the $m=0$ mode.
 \label{graphpaKerrlinecomparenote}}
\end{figure}

\begin{figure}
\epsscale{.80}
\plotone{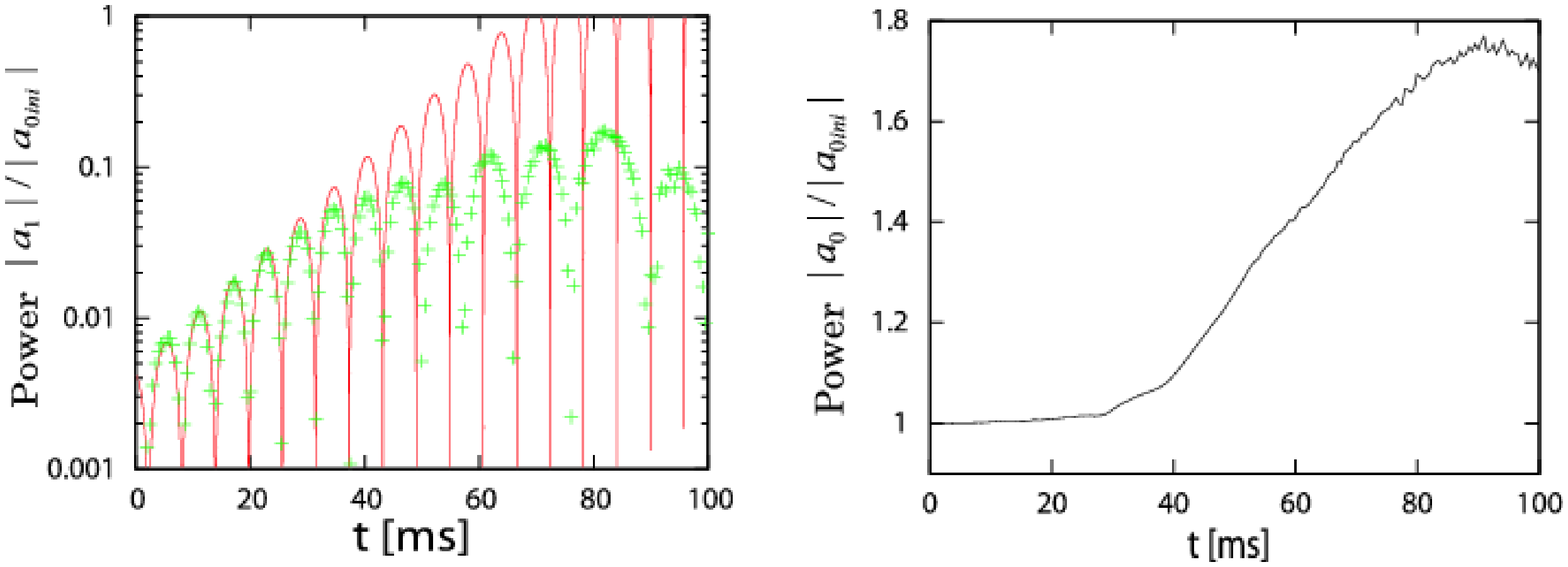}

\caption{The same as Figure~\ref{graphpaKerrlinecomparenote}
 but for Model Mk0999E4L406i.
 \label{graph999forpabhKerrnote}}
\end{figure}

\begin{figure}
\epsscale{.80}
\plotone{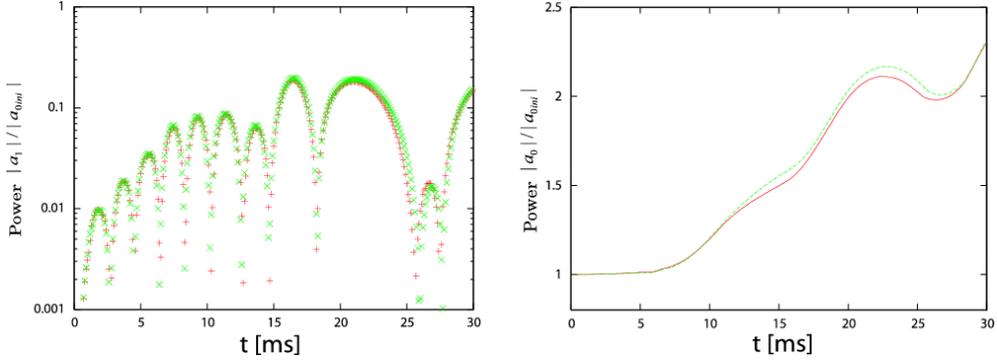}

\caption{The time evolutions of $m=1$ (left) and $m=0$ (right) modes
 for the different resolution for the standard model Mk05E4L301p.
 The red line is the result for the standard resolution in this paper,
 the green line is the result for the doubled spatial resolution.
 \label{graphconvergence}}
\end{figure}

 The oscillation timescale and growth rates obtained from
 the dynamical simulations are in good agreement with
 those found in the linear analysis
 (see the left panels of
 Figures~\ref{graphpaKerrlinecomparenote}~and~\ref{graph999forpabhKerrnote}).
 Thus, we think that the dynamical simulations are accurate enough.
 As a matter of fact,
 we have also done the simulations with doubled spatial resolutions
 both in the $r$ and $\phi$ directions.
 The computational domain is covered by $1200(r) \times 120(\phi)$ grid points
 with the smallest grid of $ \Delta r = 0.05 M_{\ast}$
 at the inner boundary.
 The difference is roughly $10\%$ at most
 (see Figure~\ref{graphconvergence}).

 It is also found from these figures that
 the results of dynamical simulations start
 to deviate from those expected by the linear analysis
 after $\sim 6$ms for Model Mk05E4L301p
 and $\sim 30$ms for Model Mk0999E4L406i,
 which indicates the beginning of the non-linear phase.
 In fact, in the right panels of these figures,
 which show the evolution of the axisymmetric $m=0$ mode,
 we find that it starts to grow
 from the time at which the non-linear phase begins.
 It is important to point out that
 the duration of the linear phase is much longer
 in Model Mk0999E4L406i than in Model Mk05E4L301p,
 since the former is more stable than the latter.

\begin{figure}
\epsscale{1.0}
\plotone{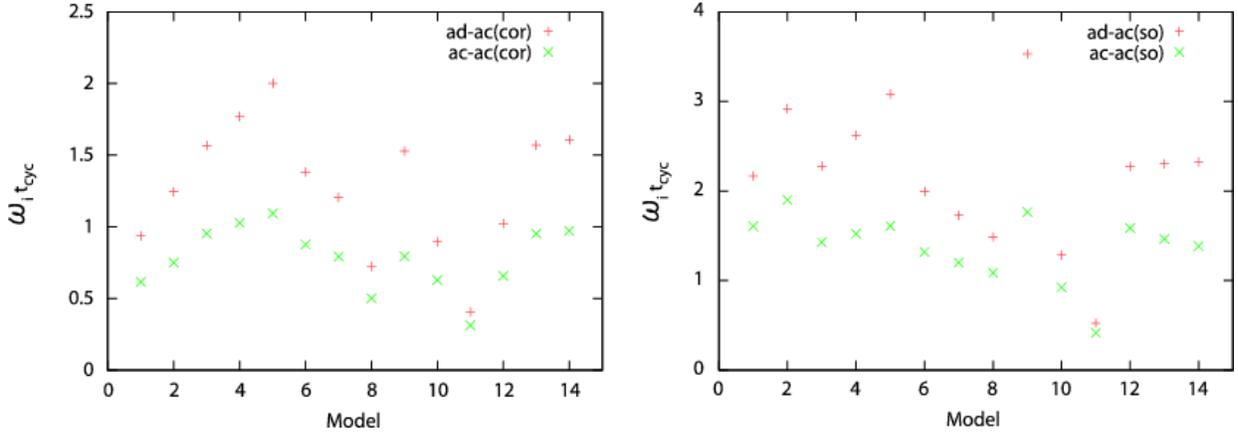}
\caption{The ratio of the growth rate to
 the frequencies of advective-acoustic (+) and
 acoustic-acoustic cycle ($\times$) for all the models.
 In the left (right) panel we assume that the corotation (inner sonic)
 point is the inner reflection point.
 See also Figure 10 in NYv1.
 \label{cycleKerr1note}}
\end{figure}

 Finally, we compare the obtained growth rates with
 the acoustic-acoustic and the advective-acoustic cycles
 between the shock surface and the inner reflection point.
 This is done to shed light on the instability mechanism.
 In this analysis, we assume that the inner reflection point is either
 the corotation radius or the inner sonic point.
 The procedure in the comparison is completely
 the same as that used in NYv1
 (see Section 5.5 in NYv1).
 The ratios of the growth rate to the period of each cycle
 are presented in Figure~\ref{cycleKerr1note}.
 It is observed that the instability is more likely to be
 of the Papaloizou-Pringle type.
 Note, however, that in general the wavelengths of acoustic perturbations
 are longer than the scale heights
 (see Table~\ref{instapropenote}) and
 we can not clearly identify the reflection point, just as in NYv1.

\section{Summary and Discussions}

 This paper is an extension of our previous work NYv1.
 The non-axisymmetric standing accretion shock instability
 around a Kerr black hole has been investigated
 by performing both linear analysis and dynamical simulations
 with fully general relativistic hydrodynamical treatments.
 This allows us not only to make clear the dependence of SASI
 on the Kerr parameter, but also to elucidate
 the effects specific to the Kerr black hole,
 such as the frame dragging.

 We have explored the range of the injection parameters
 that allows the existence of a standing shock wave in
 axisymmetric steady accretion flows
 and made clear its dependence on the Kerr parameter.
 In particular, we have demonstrated for
 large Kerr parameters and small Bernoulli constant
 in the corotation case,
 that the maximum specific angular momentum
 for the existence of standing shock wave is determined
 by {\it the limit for the multiple sonic points}.
 This is qualitatively different from the limit obtained from
 the pseudo-Newtonian analysis \citep{fog2003}.

 According to the linear analysis,
 the shock wave becomes unstable against non-axisymmetric perturbations,
 and the frame-dragging effects are also evident.
 The growth rate of the instability varies widely with the Kerr parameter.
 Since the inner transonic accretion flow
 is strongly affected by the frame-dragging
 from Kerr black hole,
 the specific angular momentum of matter
 near the black hole gets effectively increases.
 If the shock radius is unchanged,
 the intrinsic specific angular momentum of matter becomes smaller
 as the Kerr parameter gets larger.
 As a result, the Mach number of the pre-shock flow
 becomes larger (smaller) with the increasing Kerr parameter
 for the corotation (counter-rotation) flow.
 We have also investigated the dependence of the instability on
 the Kerr parameter for the same injection parameters
 instead of the same shock radius
 in the corotation case.
 The location of standing shock wave
 is shifted outward as the Kerr parameter increases,
 since the inner transonic flow gains effectively
 the specific angular momentum
 from the frame-dragging by the black hole,
 and the centrifugal force is enhanced.
 We have found that the oscillation frequency of the non-axisymmetric
 shock wave depends only on the shock radius,
 irrespective of the injection parameters
 as well as the Kerr parameters.
 At the same time,
 we find a correlation between $\omega_{r}$
 for the fundamental unstable mode and the shock radius.
 The oscillation frequency decreases monotonically
 with increasing shock radius.
 Thus, we conclude that
 the shock radius is the unique factor to decide
 the oscillation frequency.

 In the dynamical simulations,
 we have observed that the essential evolution of the perturbed system
 does not change so much from that for the Schwarzschild spacetimes.
 We have seen that the quasi-steady configuration of $m=1$ or $2$ nature
 are formed eventually for most of the models, which were also observed
 in NYv1.
 The maximum saturation level of average shock radius has been found to be
 correlated with the Mach number of the pre-shock flow,
 which is correlated in turn with the eigen-frequency
 obtained from linear analysis.
 Hence, the dependence of the saturation level in the non-linear phase
 on the Kerr parameter
 can be understood by the result of linear analysis.

 It is important to point out that
 although the ranges of the injection parameters
 that allows the existence of standing shock wave is not very wide,
 they are obtained indeed in recent progenitor models \citep{her05}.
 In fact, if we assume that the black hole mass is $M_{\ast}=3M_{\sun}$
 and the temperature of matter is less than $1$~MeV,
 the range of specific angular momentum for the existence
 of shock wave is about $\lambda_{cgs}
 \sim 3-5 \times 10^{16}$~cm$^2$/s.
 Since the standing shock wave is very robust,
 we believe that a shock wave is formed one way or another
 provided the injection parameters are appropriate.
 If such an accretion flow produces GRB somehow,
 the SASI will be a natural source of fluctuations in the GRB jets.

 In order to apply the present results to
 realistic astrophysical phenomena, such as GRB and QPO,
 there are many improvements remaining to be made.
 The radiation, viscosity and magnetic field
 should be investigated in addition to
 making the model more self-consistent,
 all of which are currently under way.
 Moreover, needless to say,
 the structure in the meridian section should be taken into account,
 which has been a major challenge for many years,
 since the accretion flow on to the black hole is inevitably
 transonic and it is difficult to impose a regularity condition
 at the sonic surface.
 In this respect, it is encouraging that the problem was
 solved by the linear perturbation method \citep{Bes2002}.
 It should be noted, however,
 the unperturbed flow is assumed to be spherically symmetric
 and the specific angular momentum and the frame-dragging
 by the Kerr black hole are small in the paper. 
 We will need to obtain steady solutions
 for wider injection and Kerr parameters,
 which is our future work.

\acknowledgments
This work was partially supported by the Grant-in-Aid for the 21st century
COE program "Holistic Research and Education Center for Physics of
Self-organizing Systems" of Waseda University and for Scientific Research
of the Ministry of Education, 
Science, Sports and Culture of Japan (17540267, 14079202).

\clearpage

\clearpage

\begin{deluxetable}{ccccccccc}
\tabletypesize{\scriptsize}
\rotate
\tablecaption{Model Parameters\label{initabnote}}
\tablewidth{0pt}
\startdata
\hline\hline
   & Kerr Parameter & Bernoulli & Specific Angular
   & Inner Sonic & Shock Point & Mach Number \\
Model & $a [M_{\ast}]$ & Constant \, $E$  & Momentum $\lambda$ $[M_{\ast}]$
   & Point $r_{inso} [M_{\ast}]$
 & $r_{sh}[M_{\ast}] $ \\
\hline
Mk00E4L343p   & 0     & 1.004 &  3.43 &  5.3 & 16.1 & 2.444 \\
Mk05E4L301p   & 0.5   & 1.004 &  3.01 &  3.6 & 16.0 & 2.682 \\
\hline
Mk00E4L350p   & 0     & 1.004 &  3.50 &  5.0 & 34.8 & 2.078 \\
Mk03E4L327p   & 0.3   & 1.004 &  3.27 &  4.1 & 35.0 & 2.137 \\
Mk06E4L297p   & 0.6   & 1.004 &  2.97 &  3.1 & 35.0 & 2.208 \\
Mk03E4L370i   & 0.3   & 1.004 & -3.70 &  5.9 & 35.0 & 2.019 \\
Mk06E4L386i   & 0.6   & 1.004 & -3.86 &  6.8 & 35.0 & 1.967 \\
Mk0999E4L406i & 0.999 & 1.004 & -4.06 &  7.9 & 35.0 & 1.904 \\
Mk01E1L350p   & 0.1   & 1.001 &  3.50 &  4.5 & 35.0 & 3.426 \\
Mk01E6L350i   & 0.1   & 1.006 & -3.50 &  5.6 & 35.0 & 1.632 \\
Mk02E8L350i   & 0.2   & 1.008 & -3.50 &  6.1 & 35.0 & 1.395 \\
\hline
Mk00E4L345p   & 0     & 1.004 &  3.45 &  5.2 & 20.8 & 2.355 \\
Mk005E4L345p  & 0.05  & 1.004 &  3.45 &  4.9 & 29.9 & 2.180 \\
Mk01E4L345p   & 0.1   & 1.004 &  3.45 &  4.6 & 44.8 & 1.931

\enddata
\tablecomments{Parameters for the initial axisymmetric steady flow
 with a standing shock wave.
 $M_{\ast}$ is the mass of the central black hole.}
\end{deluxetable}


\begin{deluxetable}{cccccccc}
\tabletypesize{\scriptsize}
\rotate
\tablecaption{Properties of Instability \label{instapropenote}}
\tablewidth{0pt}
\startdata
\hline\hline
   & maximum amplitude & Oscillation Period
   & Growth time & $|\omega_{r(f)}|$ for
   & Wavelength of Acoustic Perturbations \\
Model & of $m=0$ mode & $t_{osci}$
   & $t_{grow}/2\pi$ & fundamental node & $\lambda_{w}$ \\
\hline
Mk00E4L343p   & 3.9  & 3.7ms  & 3.2ms  & 1677 & 142.6km (31.7$M_{\ast}$) \\
Mk05E4L301p   & 9.7  & 3.7ms  & 2.5ms  & 1681 & 145.8km (32.9$M_{\ast}$) \\
\hline
Mk00E4L350p   & 3.2  & 5.9ms  & 8.2ms  & 525  & 171.9km (38.2$M_{\ast}$) \\
Mk03E4L327p   & 4.5  & 6.1ms  & 7.5ms  & 517  & 174.6km (39.4$M_{\ast}$) \\
Mk06E4L297p   & 5.5  & 6.2ms  & 6.8ms  & 503  & 181.3km (40.9$M_{\ast}$) \\
Mk03E4L370i   & 3.3  & 5.9ms  & 9.3ms  & 531  & 258.0km (58.2$M_{\ast}$) \\
Mk06E4L386i   & 2.2  & 5.9ms  & 10.7ms & 535  & 244.8km (55.3$M_{\ast}$) \\
Mk0999E4L406i & 1.8  & 11.7ms & 12.4ms & 537  & 458.3km (103.4$M_{\ast}$)\\
Mk01E1L350p   & 5.5  & 12.5ms & 6.0ms  & 504  & 343.7km (77.6$M_{\ast}$) \\
Mk01E6L350i   & 1.3  & 6.0ms  & 14.1ms & 526  & 279.1km (63.0$M_{\ast}$) \\
Mk02E8L350i   & 1.02 & 4.3ms  & 35.7ms & 511  & 202.2km (45.6$M_{\ast}$) \\
\hline
Mk00E4L345p   & 4.5  & 5.4ms  & 4.2ms  & 1156 & 186.5km (42.1$M_{\ast}$) \\
Mk005E4L345p  & 5.6  & 4.77ms & 6.7ms  & 667  & 143.9km (32.5$M_{\ast}$) \\
Mk01E4L345p   & 3.8  & 8.67ms & 11.4ms & 357  & 225.2km (50.8$M_{\ast}$)
\enddata
\tablecomments{ $t_{osci}$, $t_{grow}/2\pi$ and $\lambda_{w}$ represent
 the oscillation period,
 growth time and wavelength of acoustic perturbations, respectively,
 which are obtained by linear analysis.
 We refer the reader to the NYv1 for
 the mathematical definitions of these values.
 $\omega_{r(f)}$ is the real part of unstable fundamental eigen-frequency
 for each model.}
\end{deluxetable}

\end{document}